\documentclass[12pt,twoside]{iopart}

\usepackage[bottom]{footmisc} 
\DefineFNsymbols{rjnumber}{{*}{^3}{^4}{^5}{^6}{^7}{^8}{^9}
{^{10}}{^{11}}{^{12}}{^{13}}{^{14}}{^{15}}{^{16}}{^{17}}{^{18}}{^{19}}
{^{20}}{^{21}}{^{22}}{^{23}}{^{24}}{^{25}}{^{26}}{^{27}}{^{28}}{^{29}}
{^{30}}{^{31}}{^{32}}{^{33}}{^{34}}{^{35}}{^{36}}{^{37}}{^{38}}{^{39}}{^{40}}} 
\setfnsymbol{rjnumber} 
\usepackage{psfrag}
\usepackage{epsfig}
\newcommand{\eq}[1]{(\ref{#1})}
\newcommand{\fig}[1]{figure \ref{#1}}
\newcommand{\script}[1]{{\textrm{\scriptsize #1}}}

\newcommand{\fat}[1]{\mbox{\boldmath$#1$}}

\begin{document}
\title{Generalizing Optical Geometry}%

\author{Rickard Jonsson$^1$ and Hans Westman$^2$ \\[2mm]
{\small $^1$\it Department of Theoretical Physics, Chalmers
  University of Technology, 41296 G\"oteborg,Sweden. E-mail: rico@fy.chalmers.se}\\
{\small $^2$\it Perimeter Institute for Theoretical Physics, 31
Caroline Street North, Waterloo, Ontario N2L 2Y5, Canada. E-mail: hwestman@perimeterinstitute.ca}
\\[2mm]
\small{\rm
Submitted 2004-12-10, Published 2005-12-08\\
	 Journal Reference: Class. Quantum Grav. {\bf 23} 61
}
}
\begin{abstract}
We show that by employing the standard projected curvature as a
measure of spatial curvature, we can
make a certain generalization of optical geometry 
(Abramowicz and Lasota 1997 {\it Class. Quantum Grav.} {\bf 14} A23).
This generalization  applies to any spacetime that admits a hypersurface orthogonal
 shearfree congruence of worldlines. This is a somewhat larger class
 of spacetimes than the conformally static spacetimes assumed in
 standard optical geometry. In the generalized optical
 geometry, which in the generic case is time dependent, photons move with
 unit speed along spatial geodesics and the sideways force
 experienced by a particle following a spatially straight line is
 independent of the velocity. Also gyroscopes moving along
 spatial geodesics do not precess (relative to the forward
 direction). Gyroscopes that follow a curved spatial trajectory
 precess according to a very simple law of three-rotation. We also present an inertial force
 formalism in coordinate representation for this generalization
Furthermore, we show that by employing a new sense of spatial curvature
(Jonsson 2006 {\it Class. Quantum Grav.}  {\bf 23} 1)
closely connected to Fermat's principle, 
we can make a more extensive generalization of optical geometry that
applies to arbitrary spacetimes. In general this optical
geometry will be time dependent, but still geodesic photons move with
unit speed and follow lines that
are spatially straight in the new sense. Also, the sideways experienced
(comoving) force on a test particle following a line that is straight
in the new sense will be independent of the velocity.
\end{abstract}

\section{Introduction}
General Relativity is a theory about curved spacetime. Nevertheless
we can often gain insight by splitting spacetime into space and time
\cite{Centpar}. 
To do this we may introduce a foliation of spacetime into spatial
hypersurfaces, henceforth referred to as time slices, 
and study the geometry on these slices. In general this spatial
geometry will be time dependent, where time is a parameter
that we associate with the different slices.

For the particular case of conformally static spacetimes, Abramowicz
et. al. \cite{carter}-\cite{seb}, have demonstrated that it can be fruitful
to study a certain rescaled version of the spatial geometry, known as
the optical geometry. This rescaled geometry (which is static) has 
several features that in general the non-rescaled spatial geometry lacks: 

\begin{itemize}
\item A photon moves with unit speed (with respect to the preferred
time coordinate).

\item A photon corresponding to a spacetime null geodesic follows a spatial geodesic.

\item An observer following a spatial geodesic will experience an
acceleration (force), perpendicular to the direction of motion, that is
independent of the velocity.

\item A gyroscope following a spatial geodesic will not precess
relative to the forward direction of motion.

\end{itemize}
The optical geometry allows us to explain, in a pictorial manner,
several interesting features of black holes \cite{Centpar}. 
For example a gyroscope orbiting close to a black hole will
precess (relative to the forward direction of motion) in
the opposite direction from that of a gyroscope in orbit far away from
the hole. Another feature is that a rocket orbiting the black hole on a circle near the horizon will
require a higher outwards directed rocket thrust the faster it orbits
the hole, contrary to the situation far from the hole.

The optical geometry, with the above features, has however only
been successfully constructed in conformally static spacetimes. The
question then arises if one can generalize it to incorporate a larger class of
spacetimes. We will show that this is indeed possible. 

This article rests in part on the results of papers \cite{rickinert} and
\cite{formalgyro}. Where appropriate we will briefly review the
necessary formalism of those papers.

\subsection{The basic notation}\label{notation}
In a general spacetime, consider a reference congruence
of timelike worldlines of four-velocity $\eta^\mu$. 
We can split the four-velocity $v^\mu$ of a test particle into a part
parallel to $\eta^\mu$ and a part orthogonal to $\eta^\mu$
\begin{eqnarray}
v^\mu=\gamma(\eta^\mu + v t^\mu)
.
\end{eqnarray}
Here $v$ is the speed of the test particle relative to the congruence
and $\gamma$ is the corresponding $\gamma$-factor. The vector $t^\mu$ is a
normalized spatial vector (orthogonal to
$\eta^\mu$), pointing in the (spatial) direction of motion. 

We will also use the kinematical invariants of the congruence defined for a timelike
vector field $\eta^\mu$ as
\cite{gravitation}%
\footnote{
As concerns $\omega_{\mu \nu}$ and $\sigma_{\mu\nu}$, the
projection operators in the definition given in \cite{gravitation}
enters in a slightly different way than as presented here. For
normalized vector fields $\eta^\mu$ the definitions are equivalent,
but the form presented here is more useful if the vector field in
question is not normalized. Note also that the sign on $\omega_{\mu
  \nu}$ is a matter of convention.}
\begin{eqnarray}
a_\mu&=&\eta^\alpha \nabla_\alpha \eta_\mu    \label{katt1} \\
\theta&=&\nabla_\alpha \eta^\alpha\\
\sigma_{\mu \nu}&=&\frac{1}{2}  {P^\rho}_\nu {P^\kappa}_\mu \left(\nabla_\rho \eta_\kappa  +
\nabla_\kappa \eta_\rho \right) -\frac{1}{3} \theta P_{\mu \nu}\label{sidef}\\
\omega_{\mu \nu}&=&\frac{1}{2} {P^\rho}_\nu {P^\kappa}_\mu \left( \nabla_\rho \eta_\kappa  -
\nabla_\kappa \eta_\rho \right) \label{omdef}.
\end{eqnarray}
In order of appearance these objects denote the acceleration
vector, the
expansion scalar, the shear tensor and the rotation tensor. 
We will also employ what we may denote the expansion-shear tensor
\begin{eqnarray}
\theta_{\mu \nu}=\frac{1}{2} {P^\rho}_\nu {P^\kappa}_\mu \left(\nabla_\rho \eta_\kappa  +
\nabla_\kappa \eta_\rho \right) .
\end{eqnarray}
Throughout the article we will use $c=1$ and adopt the spatial sign convention
$(-,+,+,+)$. The projection operator\footnote{Applying this operator
to a vector extracts the part of the vector that is
  orthogonal to $\eta^\mu$.
} 
along the congruence  then takes the form ${P^\alpha}_\beta\equiv
{\delta^\alpha}_\beta+\eta^\alpha \eta_\beta$. Vectors that are
orthogonal
 to $\eta^\mu$ , will be referred to as {\it spatial} vectors.
We will also find it convenient to introduce the suffix $\perp$. When applied to a
four-vector, like $[K^\mu]_\perp$, it selects
the part within the brackets that is perpendicular to both $\eta^\mu$
and $t^\mu$\footnote{Hence $[K^\mu]_\perp={P^\mu}_\alpha
  (K^\alpha-K^\beta t_\beta t^\alpha)$.}.

\section{Generalizing the optical geometry}\label{generalize}
A key feature of optical geometry is that we have a {\it space} that
accounts for the motion of geodesic photons. To define a space, given
a spacetime, we specify a congruence of timelike worldlines
(generated by a normalized vector field $\eta^\mu$). Every worldline corresponds to a single point in the space. 

To completely account for the behavior of photons we need also to
introduce a (global) time in which the position of photons in the
spatial geometry is evolved. This is done by introducing time
slices. These time slices can be defined by a single function
$t(x^\mu)$ (where the slices are defined by $t=const$)%
\footnote{We can introduce spacelike time slices within a finite
  region around any point in an arbitrary spacetime. In globally hyperbolic spacetimes such
 slices can be
  globally defined.}.

It is easy to realize that if we want the velocity%
\footnote{We have not introduced a three-metric yet so we cannot
really talk about velocities. Still given two nearby congruence lines
A and B, we can compare the coordinate time ($t$) it takes a photon to
move from A to B and vice versa. If the time slice is not orthogonal
to the congruence then $t_{AB}$ will not in general equal $t_{BA}$ for
infinitesimally displaced congruence lines. Thus the velocity would be
dependent on the spatial direction of motion.}
of photons to be independent of direction we must have the time slices orthogonal to the congruence. 
This means that, given $t(x^\mu)$, the local direction of the congruence is uniquely determined by
\begin{eqnarray} \label{bat}
\eta_\mu=-e^{\Phi}   \nabla_\mu t
.
\end{eqnarray}
The function $\Phi$ can be determined by demanding $\eta^\mu
 \eta_\mu=-1$. Then we can (in principle) integrate \eq{bat} to find
 the congruence lines uniquely. 

Photon geodesics are invariant under conformal rescalings
of the metric. Thus without affecting the spacetime properties with
respect to geodesic photons, we can rescale the metric around every spacetime
point with a factor $e^{-2\Phi}$. Letting a tilde denote objects
related to the rescaled spacetime, we have $\tilde{g}_{\mu
\nu}=e^{-2\Phi} g_{\mu \nu}$.  The conformal
rescaling effectively removes time dilation (lapse) so that
$dt=d\tilde{\tau}_0$, where $d\tilde{\tau}_0$ is the proper time along
the congruence in the rescaled spacetime.
Relative to the rescaled space, photons move with unit speed
$\frac{d\tilde{s}}{dt}=1$. Thus the first point in the list of
features in the introduction (photons move with constant speed with
respect to coordinate time), we can always achieve. 
What about the second point? Under what conditions will
photons follow straight spatial lines, and indeed what do we mean by
following a straight line if the spatial geometry is time dependent?

\section{Generalizing the optical geometry using the projected curvature}\label{projgen}
As regards what is spatially straight, most likely the first thing that comes to mind (at least it was
for the authors of this article), is to consider a projection of the
null trajectory in question down along the congruence to the local
slice%
\footnote{If the congruence has no rotation there exists a finite
sized slicing orthogonal to the congruence. If the congruence is
rotating we can still introduce a slicing that is orthogonal at the
point in question. It is easy to realize that whatever such locally
orthogonal slicing we
choose, the projected curvature and curvature directions will be the same, and are thus well defined.}%
. If the spatial
curvature of the projected curve vanishes -- then we say that the trajectory is straight. In \cite{rickinert} a general formalism of
inertial forces in terms of the projected curvature is derived using
an arbitrary congruence of timelike worldlines. The (projected)
four-acceleration of the test particle in question can be decomposed
as (see section \ref{notation} concerning notation)
\begin{eqnarray}\label{rattok}
\frac{1}{\gamma^2}{P^\mu}_\alpha \frac{Dv^\alpha}{D\tau}
= a^\mu 
+ 2v \left[t^\alpha \nabla_\alpha \eta^\mu \right]_\perp +  v t^\mu t^\alpha
  t^\rho\nabla_\rho \eta_\alpha  + \gamma
\frac{dv}{d\tau} t^\mu + v^2 \frac{n^\mu}{R}.
\end{eqnarray}
Here $R$ is the projected spatial curvature radius and $n^\mu$ is a 
normalized spatial vector (orthogonal to both $t^\mu$ and $\eta^\mu$),
  pointing in the direction of projected spatial curvature 
(the principal normal).
The left hand side of \eq{rattok} can be expressed in terms of the forces acting
on the test particle. We have \cite{rickinert}
\begin{eqnarray}\label{finapp}
{P^\mu}_\alpha \frac{D v^\alpha}{D\tau}&=&\frac{1}{m}  {P^\mu}_\alpha
f^\alpha =\frac{1}{m}\left(\gamma F_{\parallel} t^\mu + F_{\perp} m^\mu\right).
\end{eqnarray}
Here $m^\mu$ is a normalized spatial vector orthogonal to both $t^\mu$
and $\eta^\mu$. $F_\perp$ and $F_\parallel$ are respectively the forces perpendicular and parallel to
the direction of motion, as experienced in a system comoving with the
test particle in question. The right hand side of \eq{rattok} can be expressed in terms of the
kinematical invariants of the congruence, through the identity 
\cite{gravitation} $\nabla_\nu \eta_\mu=\omega_{\mu \nu}+\theta_{\mu
  \nu}-a_\mu \eta_\nu$. Then \eq{rattok} takes the form \cite{rickinert}
\begin{eqnarray}\label{hutt}
\frac{1}{m \gamma^2}\left(\gamma F_{\parallel} t^\mu +  F_{\perp} m^\mu\right)
= &&{\hspace{-0mm}}a^\mu  
+2 v \left[t^\beta ( {\omega^\mu}_\beta+{\theta^\mu}_\beta)\right]_\perp
+v t^\alpha t^\beta \theta_{\alpha \beta} t^\mu \\&&{\hspace{-0mm}} \nonumber
+ \gamma \frac{dv}{d\tau} t^\mu + v^2 \frac{n^\mu}{R}
.
\end{eqnarray}
On the right hand side we have
first three terms that enter as inertial forces (if we multiply them
by $-m$), and the last two terms describe the motion (acceleration) relative to
the reference congruence. 

Following \cite{rickinert} one can form a corresponding rescaled version of \eq{rattok} by
  putting a tilde on everything in \eq{rattok}. Next one finds the
  general relation between rescaled and non-rescaled
  four-acceleration, the result is given by given by
  \eq{fouracc2}. Setting
  $\tilde{a}^\mu=0$ and $\tilde{\omega}_{\mu \nu}=0$ as is appropriate for the
  congruence and rescaling at hand, also using
  $\tilde{t}^\mu=e^\Phi t^\mu$ and $\tilde{m}^\mu= e^\Phi m^\mu$ and using \eq{finapp} one
  readily gets \cite{rickinert}
\begin{eqnarray}\label{much1c}
\frac{1}{m\gamma^2} e^\Phi \left(
\frac{F_\parallel}{\gamma} \tilde{t}^\mu + F_\perp \tilde{m}^\mu
\right)
=&&
\frac{1}{\gamma^2} \tilde{P}^{\mu \rho}
\tilde{\nabla}_\rho \Phi +
\frac{v}{\gamma^2}(\tilde{\eta}^\rho\tilde{\nabla}_\rho \Phi   + \tilde{t}^\alpha \tilde{t}^\beta \tilde{\theta}_{\alpha \beta}   )
\tilde{t}^\mu  \\\nonumber
&&+2 v \left[\tilde{t}^\beta {\tilde{\theta}^\mu}{}_\beta \right]_\perp
+  \frac{dv}{d\tilde{\tau}_0} \tilde{t}^\mu + v^2 \frac{\tilde{n}^\mu}{\tilde{R}}
.
\end{eqnarray}
Recall that a tilde implies that the object is related to the
rescaled spacetime. As concerns $\gamma$ and $v$ we have however omitted
the tilde since these are the same as their non-rescaled analogues.
Note also that $F_\parallel$ and $F_\perp$ are the real (non-rescaled)
comoving forces. For a discussion of how to interpret this expression
in terms of inertial forces we refer to \cite{rickinert}, and the
discussion in section \ref{disk} of this paper. 

One sees from \eq{much1c} (set the left hand side to zero and $v=1$) that the projected optical curvature of a
geodesic photon vanishes, for all spatial directions, if and only if
$[\tilde{t}^\beta {\tilde{\theta}^\mu}{}_\beta]_\perp=0$ for all
$\tilde{t}^\mu$. We also readily see that the sideways (perpendicular)
experienced force%
\footnote{Notice that the force that we are referring to here is the
force as received in a system comoving with the test-particle. If we
on the other hand consider the reference congruence observers to be
providing the sideways force, this force is in fact smaller than the
received force by a $\gamma$-factor and is hence not independent of
the velocity, see section \ref{fnote}.}
$F_\perp$ is independent of the velocity when following a
trajectory whose projected curvature vanishes, if and only if
$[\tilde{t}^\beta {\tilde{\theta}^\mu}{}_\beta]_\perp=0$%
\footnote{For this case the sideways force is given by 
$F_\perp=m e^{-\Phi} \sqrt{\tilde{g}_{\mu \nu} [\tilde{P}^{\mu
\rho} \tilde{\nabla}_\rho \Phi   ]_\perp [\tilde{P}^{\nu
\beta} \tilde{\nabla}_\beta \Phi    ]_\perp    }$.}. 
As discussed in \cite{rickinert}, and reviewed in \ref{shearl}, this holds for all
directions $\tilde{t}^\mu$ if and only if the congruence is {\it
shearfree}%
\footnote{The shear tensor in the rescaled spacetime is given by
$\tilde{\sigma}_{\mu \nu}= e^{-\Phi} \sigma_{\mu \nu}$, so the shear
tensor in the rescaled spacetime vanishes if and only if it does so in
the non-rescaled spacetime.}.
Note that for the standard static optical geometry,  we have
$\tilde{\theta}_{\mu \nu}=0$, thus the congruence is trivially
shearfree. Incidentally, for this case \eq{much1c} can be written as 
\begin{eqnarray}\label{much1cred}
\frac{e^\Phi}{m}  \left(
\frac{F_\parallel}{\gamma} \tilde{t}^\mu + F_\perp \tilde{m}^\mu
\right)
=&&
\tilde{P}^{\mu \rho}
\tilde{\nabla}_\rho \Phi +
v(\tilde{\eta}^\rho\tilde{\nabla}_\rho \Phi   )
\tilde{t}^\mu  +  \gamma^2 \frac{dv}{d\tilde{\tau}_0} \tilde{t}^\mu +
\gamma^2 v^2 \frac{\tilde{n}^\mu}{\tilde{R}}
.
\end{eqnarray}
We conclude that for shearfree congruences we can always manage the first three points of
the list of optical geometry features given in the introduction. 
Now what what about the fourth point, concerning gyroscope precession?

\subsection{Gyroscope precession}\label{spinproj}
The spin vector $S^\mu$ of an ideal gyroscope
transported without any torque acting on it in a comoving system,
along a trajectory of four-velocity $v^\mu$, obeys the
Fermi-Walker equation
\begin{eqnarray}
\frac{D S^\mu}{D\tau} = v^\mu S_\alpha \frac{D v^\alpha}{D\tau}
.
\end{eqnarray}
Introducing $\tilde{S}^\mu=e^\Phi S^\mu$ and $\tilde{v}^\mu=e^\Phi v^\mu$, using the orthogonality of
$S^\mu$ and $v^\mu$, it is a quick exercise (carried out in \ref{fermiapp}) to show that this implies
\begin{eqnarray}
\frac{\tilde{D} \tilde{S}^\mu}{\tilde{D} \tilde{\tau}} = \tilde{v}^\mu \tilde{S}_\alpha\frac{\tilde{D} \tilde{v}^\alpha}{\tilde{D} \tilde{\tau}} 
.
\end{eqnarray}
Thus a spin vector which is Fermi-Walker transported relative to the
non-rescaled spacetime is Fermi-Walker transported also relative to the
rescaled spacetime if we just rescale the spin vector itself.
In particular, a gyroscope initially pointing in
the forward direction $t^\mu$ precesses relative to the forward direction in
the standard spacetime if and only if it does so in the rescaled spacetime.

In \cite{formalgyro}, a general formalism of gyroscope precession with
respect to a reference congruence is discussed. 
Here one considers the spin vector
that one {\it would get} if one were to momentarily stop (by a pure boost)
the gyroscope with respect to the reference congruence. This spin
vector is called the stopped spin vector, denoted by
$\bar{S}^\mu$ and related to $S^\mu$ through 
\begin{eqnarray}\label{stop1}
\bar{S}^\mu&=&\left[{\delta^\mu}_\alpha +\eta^\mu \eta_\alpha
+\left(\frac{1}{\gamma}-1\right)t^\mu t_\alpha \label{s2}\right] S^\alpha
.
\end{eqnarray}
While $\bar{S}^\mu$ is orthogonal to the
congruence it is not simply the projected part of the standard spin
vector (in general both the norm and the spatial direction of these two
objects differ). The reason for using $\bar{S}^\mu$ rather than
$S^\mu$ is that $\bar{S}^\mu$ (unlike the projected part of $S^\mu$)
obeys a simple law of (three-dimensional) rotation
\begin{eqnarray}\label{kottfinal}
\frac{D \bar{S}^\mu}{D\tau}=\frac{\gamma v
}{\gamma+1} \bar{S}^\alpha \left( t^\mu \wedge
\left[\frac{D}{D\tau}\left(v_\alpha + \eta_\alpha \right) \right]_\perp \right)
+\eta^\mu \bar{S}^\alpha \frac{D\eta_\alpha}{D\tau}
.
\end{eqnarray}
Here the last term 
insures that
orthogonality to the congruence is preserved. 
Note also that $\frac{D}{D \tau}$ 
means covariant differentiation along the gyroscope worldline.
The contraction with the wedge product (defined as $k^\mu \wedge b_\alpha \equiv
k^\mu b_\alpha - k^\mu a_\alpha$) corresponds to the three-rotation%
\footnote{
Choose inertial coordinates adapted to the the congruence so that
$\bar{S}^\mu=(0,{\bf \bar{S}})$, $t^\mu=(0,{\bf \hat{t}})$ and
$n^\mu=(0,{\bf \hat{n}})$. Then \eq{kottfinal} amounts to
$
\frac{d{\bf \bar{S}}}{d\tau}=\gamma v (\gamma-1)\left[{\bf \hat{t}}({\bf
\bar{S}} \cdot \frac{{\bf \hat{n}}}{R} ) -\frac{{\bf \hat{n}}}{R} ({\bf \bar{S}}
\cdot {\bf \hat{t}} )  \right]
$. The expression within the brackets is a vector
triple product and we may write it as a double cross
product. Letting ${\bf v}=v {\bf \hat{t}}$ we get
$\frac{d{\bf \bar{S}}}{d\tau}=\gamma 
(\gamma-1) \left(\frac{{\bf \hat{n}}}{R}\times {\bf v}\right) \times {\bf \bar{S}} 
$, which is a simple equation of three-rotation.}.
Furthermore one considers a spacetime analogue of standard spatial
transport
\begin{eqnarray}\label{ps}
\frac{Dk^\mu}{D\tau} = \gamma k^\alpha {\omega^\mu}_\alpha
 + \gamma k^\alpha ({\theta^\mu}_\beta t^\beta \wedge
 t_\alpha) 
+ \eta^\mu k^\alpha \frac{D\eta_\alpha}{D\tau} 
.
\end{eqnarray}
The transport as defined here is norm preserving, and also preserves angles
between transported vectors%
\footnote{The general idea of the transport equation is most readily
  understood considering a rigid 
(non-shearing and non-expanding)
  reference congruence. When the reference frame rotates -- so does
  the parallel transported vector.}. 
Considering a spatially straight line ($1/R=0$), and a vector
momentarily aligned with the forward direction, the transport is
defined such that the vector remains in the forward direction (analogous to
standard spatial transport).

Given \eq{kottfinal} and \eq{ps} one can form an equation for how fast
the stopped spin vector deviates from a corresponding (spatially)
parallel transported vector. For the case of a congruence with vanishing 
rotation (as is appropriate for the hypersurface orthogonal congruence we
are here considering) one finds \cite{formalgyro}
\begin{eqnarray}\label{finp}
\frac{D_\script{ps} \bar{S}^\mu}{D_\script{ps} \tau}=\bar{S}^\alpha&&\hspace{-0mm}
\Bigg[ \gamma^2 v (t^\mu \wedge a_{\alpha})  + (2\gamma^2-1) (t^\mu \wedge t^\beta
\theta_{\alpha \beta}) 
+ \nonumber\\&&\hspace{-0mm}
\hspace{2mm}+ \gamma v
(\gamma -1) \left(t^\mu \wedge \frac{n_\alpha}{R} \right)
\Bigg]
.
\end{eqnarray}
Here the suffix 'ps' is short for 'projected straight'%
\footnote{In the coming
section we will consider a different derivative connected to a
different notion of straightness, that will get the suffix 'ns'.}.
What \eq{finp} tells us is how the
stopped spin vector precesses relative to a frame that is spatially
parallel transported with respect to the reference congruence. 

We can of course also consider the analogue of \eq{finp} for the rescaled
spacetime. There the congruence acceleration vanishes and we are left with
\begin{eqnarray}\label{finpo}
\frac{\tilde{D}_\script{ps}
\tilde{\bar{S}}^\mu}{\tilde{D}_\script{ps} \tilde{\tau}}=
\tilde{\bar{S}}^\alpha
\left[(2\gamma^2-1) (\tilde{t}^\mu \wedge \tilde{t}^\beta \tilde{\theta}_{\alpha \beta}) +\gamma v
(\gamma -1) \left(\tilde{t}^\mu \wedge \frac{\tilde{n}_{\alpha}}{\tilde{R}} \right)
\right]
.
\end{eqnarray}
In particular, demanding that the gyroscope
should remain pointing in the forward direction $t^\mu$ as it follows a
spatially straight line in the rescaled spacetime, the left hand side
of \eq{finpo} must vanish and we get
\begin{eqnarray}\label{fini}
0=\tilde{t}^\alpha (\tilde{t}^\mu \wedge \tilde{t}^\beta \tilde{\theta}_{\alpha \beta})
.
\end{eqnarray}
This equation can be simplified to
\begin{eqnarray}\label{fina}
0=[\tilde{t}^\beta { \tilde{\theta}^\mu}{}_\beta]_\perp
.
\end{eqnarray}
As mentioned earlier, \eq{fina} holds for all directions $\tilde{t}^\mu$ if
and only if the congruence is shearfree.
So a gyroscope (initially directed in
the forward direction) will not precess relative to the forward
direction, when transported along a spacetime trajectory whose spatial
projection has vanishing curvature relative to the rescaled spacetime,
if and only if the congruence is shearfree. 

When the congruence is shearfree, \eq{finpo} is simplified to
\begin{eqnarray}\label{ratt}
\frac{\tilde{D}_\script{ps}
\tilde{\bar{S}}^\mu}{\tilde{D}_\script{ps}
\tilde{\tau}}=\tilde{\bar{S}}^\alpha \gamma v
(\gamma -1) \left(\tilde{t}^\mu \wedge \frac{\tilde{n}_{\alpha}}{\tilde{R}} \right)
.
\end{eqnarray}
Comparing with \eq{finp} (setting $a^\mu=0$ and $\theta^{\mu \nu}=0$
corresponding to an inertial congruence), we see that if we consider the gyroscope
precession with respect to the rescaled spacetime, there is only
standard Thomas-precession (see e.g. \cite{formalgyro,
  intuitivegyro}). 

Note that while \eq{ratt} is a four-vector relation, it is effectively a
three-dimensional equation since all the terms are orthogonal to $\eta^\mu$. Letting $(0,{\bf \tilde{v}})=v
\tilde{t}^\mu$, $(0,{\bf \tilde{\hat{n}}})=\tilde{n}^\mu$ and $(0,{\bf
  \tilde{\bar{S}}})=\tilde{\bar{S}}^\mu$ in coordinates locally comoving with the the
reference congruence we can express \eq{ratt} in manifest three-form
as (see footnote on previous page for details)
\begin{eqnarray}\label{ratt2}
\frac{D {\bf \tilde{\bar{S}}}}{D t}=(\gamma-1) 
\left( \frac{{\bf \tilde{\hat{n}}}}{\tilde{R}}\times {\bf \tilde{v}} \right)\times 
     \tilde{{\bf \bar{S}}}
\end{eqnarray}
So here we see how the (rescaled and stopped) spin vector precesses
relative to a corresponding frame that is parallel transported with 
respect to the optical geometry. Note the absence of explicit factors
$e^{\Phi}$ and how very simple this law of precession is. 

Considering gyroscope precession relative to some curved (rescaled) spatial
geometry  we must take into account that a parallel transported frame
will in general be rotated relative to its initial configuration if we transport it along some
closed spatial trajectory.
For motion in the equatorial plane of some
axisymmetric static optical geometry this can easily be dealt with by introducing a
reference frame that rotates relative to a local frame spanned by ${\bf \tilde{\hat{r}}}$ and $\fat{ \tilde{\hat{\varphi}}}$, in the
same manner as a parallel transported reference frame does on a plane.
Such a reference frame always returns to its initial configuration after
a full orbit. In \cite{formalgyro} the effective rotation relative to
this new frame of reference is derived. If the line element can be written on the form 
\begin{eqnarray}
d\tilde{s}^2=\tilde{g}_{\tilde{r} \tilde{r}} d\tilde{r}^2 + \tilde{r}^2 d\varphi^2
\end{eqnarray}
the effective rotation vector can be written as
\begin{eqnarray}\label{katt2}
\fat{\tilde{\Omega}}_{\script{effective}}=&&
(\gamma-1)\left(\frac{{\bf \tilde{\hat{n}}}}{\tilde{R}} \times {\bf \tilde{v}} \right) 
+\frac{1}{\tilde{r}} \left(\frac{\pm 1}{\sqrt{\tilde{g}_{\tilde{r} \tilde{r}}}}-1
\right) {\bf \tilde{v}} \times {\bf  \tilde{\hat{r}}} 
.
\end{eqnarray} 
We have here included a $\pm$ sign. 
If ${\bf \tilde{\hat{r}}}$, which is assumed to be pointing away from the
center of symmetry, points in the direction of increasing $\tilde{r}$
we have the positive sign, otherwise we should use the negative sign%
\footnote{For the standard optical geometry of a
  black hole there is a neck (minimum value of $\tilde{r}$) at the
  photon radius. For this geometry we should use the positive sign
  outside of the photon radius and the negative sign inside of this
  radius. Note that
  $\frac{1}{\sqrt{\tilde{g}_{\tilde{r} \tilde{r}}}}=0$ at the photon
  radius, so there is no discontinuity in $\fat{\tilde{\Omega}}_{\script{effective}}$.}.
Note that \eq{katt2}, for the particular case of motion in an
  axisymmetric spatial geometry, gives the precession relative to the 'would-be-flat'
reference frame in terms of the parameter time $t$%
\footnote{This time equals the local time of the congruence observers
  in the rescaled spacetime. In the case of standard optical
  geometry for a Schwarzschild black hole, it is simply the Schwarzschild time.}.

\subsection{Conclusion as regards the standard projected curvature}
We have seen that the optical geometry, as presented here, retains all of
the features listed in the introduction given that the {\it
shear-tensor} of the congruence in question vanishes. This corresponds to a larger set
of spacetimes than the conformally static spacetimes, see section \ref{unique}. We have also
seen that gyroscopes precess according to a very simple law of
rotation with respect to the optical geometry.

\section{Generalizing the optical geometry using a different curvature
  measure}\label{news}
While vanishing projected curvature is likely to be the first notion of spatial
straightness that comes to mind, it is perhaps not the most natural for
all cases. 
In \cite{rickinert}, a novel definition 
is proposed via a variational principle. Let $ds$ be the distance
traveled for a test particle as seen from the local congruence
observers ($ds=\gamma v d\tau$). Parameterizing the
trajectory by $\lambda$, the integrated distance $\delta s$
along a trajectory connecting two fixed spacetime points can
be written as
\begin{eqnarray}\label{sta}
\delta s&=&\int ds \\
&=& \int \sqrt{P_{\mu \nu} \frac{dx^\mu}{d\lambda}
\frac{dx^\nu}{d\lambda} } d\lambda.
\end{eqnarray}
One can show \cite{rickinert} that in order for this action to be stationary (minimized) with
respect to variations of the trajectory that are perpendicular to
$\eta^\mu$ and $t^\mu$, the projected curvature must obey
\begin{eqnarray}\label{final}
v \frac{n_\mu}{R}=-2 [t^\alpha \theta_{\alpha \mu}]_\perp
.
\end{eqnarray}	
Trajectories that are obeying this relationship are thus minimizing%
\footnote{Strictly speaking, the distance traveled with respect to the
  congruence observers is stationary with
  respect to variations perpendicular to $\eta^\mu$ and $t^\mu$ if the
  curvature obeys \eq{final}. }
the spatial distance traveled and are said to be
straight in the new sense or simply new-straight (for want of a better
  name).
Thus the two notions of straightness differ (in general) if and only if there is shear. 
Notice that the curvature relation of \eq{final} is
velocity dependent. For more details and some intuition of why the two
notions of straightness differ see \cite{rickinert}.

One may introduce a new curvature measure from how
fast a trajectory deviates from a corresponding (same $v$ and $t^\mu$)
new-straight trajectory. Denoting the corresponding curvature direction
and curvature radius by $\bar{n}^\mu$ and $\bar{R}$ respectively, the inertial
force formalism \cite{rickinert}, in an optically rescaled spacetime
(analogous to the outline in the preceeding section) takes the form
\begin{eqnarray}\label{optisk}
\frac{1}{m\gamma^2} e^\Phi \left(
\frac{F_\parallel}{\gamma} \tilde{t}^\mu + F_\perp \tilde{m}^\mu
\right)=&&\hspace{-0mm}
\frac{1}{\gamma^2} \tilde{P}^{\mu \rho} \tilde{\nabla}_\rho \Phi +
\frac{v}{\gamma^2}( \tilde{t}^\alpha \tilde{t}^\beta
\tilde{\theta}_{\alpha \beta} +    \tilde{\eta}^\rho\tilde{\nabla}_\rho
\Phi) \tilde{t}^\mu  \\&&\hspace{-0mm} \nonumber
+ \frac{dv}{d\tilde{\tau}_0} \tilde{t}^\mu + v^2 \frac{\tilde{\bar{n}}^\mu}{\tilde{\bar{R}}}
.
\end{eqnarray}
It follows immediately that a geodesic photon (set the left hand
side to $0$ and $v=1$) has zero curvature ($1/\tilde{\bar{R}}=0$) in the new sense relative to
the rescaled spacetime. Also, the sideways force on a massive
particle following a straight line is independent of the velocity. 
Notice that this holds independent of whether the congruence is
shearing or not. So in fact, with the new sense of curvature, for {\it
any} spacetime and {\it any} spacelike foliation (and corresponding rescaling),
a geodesic photon follows a spatially straight line, and the sideways
force on an object following a spatially straight line is independent
of the velocity. 

\subsection{Gyroscope precession}
In \cite{formalgyro}, a spin precession formalism connected
to the new-straight curvature measure is presented, analogous to \eq{finp} above
for the projected curvature measure. Relative to the rescaled
spacetime (set $a^\mu=0$, $\omega^{\mu \nu}=0$ and put tilde on
everything) we have from \cite{formalgyro} 
\begin{eqnarray}
\frac{\tilde{D}_\script{ns} \tilde{\bar{S}}^\mu}{\tilde{D}_\script{ns} \tilde{\tau}}=\tilde{\bar{S}}^\alpha&&\hspace{-0mm}
\left[ -(\tilde{t}^\mu \wedge \tilde{t}^\beta \tilde{\theta}_{\alpha \beta}) +\gamma v
(\gamma -1) \left(\tilde{t}^\mu \wedge \frac{\tilde{\bar{n}}_{\alpha}}{\tilde{\bar{R}}} \right)
\right]
.
\end{eqnarray}
Here 'ns' stands for new-straight%
\footnote{Note that the bar in $
\tilde{\bar{n}}_{\alpha}$ and the bar in ${\tilde{\bar{R}}}$ have nothing to do
with the bar in $\tilde{\bar{S}}^\alpha$.}.
While the term containing the expansion-shear tensor is simpler in
this equation,
compared to \eq{finp}, it is not vanishing. Thus, analogous to the discussion in section \ref{spinproj}, a
gyroscope initially directed in the forward direction will remain in
the forward direction, as we move along an arbitrary line that is straight in the new
sense relative to the rescaled spacetime, if and only if the
shear-tensor of the congruence vanishes.

\subsection{Fermat's principle and the new-straight curvature}
As discussed in \cite{rickinert}, the new-straight curvature relative
to the rescaled spacetime is
closely related to Fermat's principle. Indeed for a photon in the rescaled spacetime, the
coordinate time it takes for a photon to go from a certain event along
one spatial point (congruence line) to another spatial 
point (congruence line) is given by
\begin{eqnarray}\label{derhuygens}
\delta t=\int dt=\int d\tilde{s}
.
\end{eqnarray}  
Fermat's principle states that a null trajectory is a
geodesic if and
only if it extremizes $\delta t$%
\footnote{We here naturally refer to the part of the null
  trajectory that connects the two congruence lines in question. By extremizes we mean that we may have a minimum or a
  saddlepoint in $\delta t$ with respect to (null-preserving)
  variations around the trajectory in question.}.
Also, by {\it definition} $\int d\tilde{s}$ is
  extremized%
\footnote{Strictly speaking $\int d\tilde{s}$ is extremized
  with respect to variations perpendicular to the trajectory. For the
  case of null-preserving variations, perpendicular
  variations are however (to the necessary order) the only allowed
  variation.}
 if and only if the curvature in the new sense (with
  respect to the rescaled spacetime) vanishes
, which according to \eq{derhuygens} means that $\delta t$ is
extremized. 
It follows that any null geodesic has vanishing spatial curvature in
the new sense relative to the rescaled spacetime. 

The connection between Fermat's principle and straight lines in the
optical geometry was realized, for conformally static spacetimes, a
long time ago. With the new definition of curvature
the connection holds in any spacetime.

\subsection{Conclusion regarding the new sense of curvature in
relation to optical geometry}
We have seen that with the new sense of curvature, photons move along
spatial geodesics for any slicing and corresponding rescaling in any
spacetime. Also the sideways force on a particle following a spatially
straight line will be independent of the spatial velocity. Gyroscopes
following straight lines will however precess (in general) relative to
the forward direction.

\section{Some comments on rescalings and other transformations}
\begin{figure}[ht]
  \begin{center}
	\psfrag{Rightmoving photon}{Right-moving photon}
	\psfrag{Leftmoving photon}{Left-moving photon}
	\psfrag{Spatial Geodesic}{Spatial geodesic}
      	\epsfig{figure=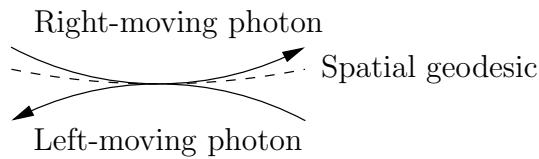,width=7cm,angle=0}
      	\caption{Photons moving in opposite spatial directions will in general have different projected
      	curvatures when the congruence is shearing or rotating. 
        This means that the motion of photons cannot
      	generally correspond to projected geodesics of any three-geometry for a
      	rotating or shearing congruence.} 
     	\label{fig1}
  \end{center} 
\end{figure}

We have in this article used conformal rescalings more or less
without motivation.  In principle one can imagine other
transformations. In particular we can consider a transformation of just the spatial
geometry, and do this in such a
manner that no matter what shear or rotation the reference
congruence has, all the projected geodesic photons get vanishing spatial curvature. 
But in fact, this is not generally doable. To see this, consider the projection of a left-moving and a
right-moving geodesic photon when the congruence
rotates (think in 2+1 inertial coordinates). The projected
curvature direction  (relative to the standard on-slice
geometry) will be {\it opposite} for the
two directions as illustrated in \fig{fig1}. 

The two projected trajectories are aligned at the origin,  but are then
separating. A geodesic aligned with the two trajectories can however 
not split in two. Hence there simply exists {\it no} three-geometry
relative to which the projections of general geodesic photons will
correspond to spatial geodesics if the congruence is rotating. The
same argument applies when the congruence is shearing.
So whatever transformation of the spatial metric we are considering, 
using the standard
projected curvature, the optical congruence would have to be
rotationfree and shearfree for geodesic photons to follow spatial geodesics.

But why conformal rescalings? Well, as we argued before, 
we need a congruence and a corresponding
orthogonal slicing to get an isotropic speed of light (with respect to
coordinate time), no matter what spatial metric we come up with. Also, for the chosen labeling of the
time-slices, if we want the speed of light (with respect to the
transformed) spatial metric to be unit everywhere ($\frac{d \tilde{s}}{dt}=1$), where
$d\tilde{s}$ is the transformed spatial distance (no matter what
transformation we are considering), then a rescaling of spatial
distances by a factor $e^{-\Phi}$ is in fact the only option. 
That geodesic photons follow projected straight lines relative to the
rescaled space, we may see as a pure bonus when there is no shear. 
Notice however that even if we would relax the unit
speed requirement, we would still need a shearfree congruence to get
photons to follow straight spatial lines in the projected sense.

In summary, as concerns the unit speed requirement, the rescaling is the
only option. Concerning the requirement that the projection of a
null geodesic should correspond to a spatially straight line,
no other transformation would make a better job%
\footnote{In other words
no other transformation would be applicable to a larger set
of spacetimes (and congruences) than those already considered for the
generalized (in the projected sense) optical geometry.}. 
Also, using the
new-straight curvature, we get {\it all} of the features concerning
photons that we want using the rescaling scheme. 

Notice that (as we have argued above) the fact that photon
geodesics are unaffected by the conformal rescaling, is not of any
{\it direct} importance. Considering the projected curvature, we can in principle imagine some other spacetime
transformation that does affect null geodesics, but that gives a vanishing projected curvature (in the
transformed spacetime) for a photon that is a geodesic in the original spacetime.
However, for conformally static
spacetimes, assuming that we rescale away time dilation (lapse), the fact
that photon geodesics are unaffected by the rescaling {\it trivially} means
that the projected spatial curvature of a geodesic photon must
vanish%
\footnote{Thinking of geodesics as maximizing the proper time (null
  geodesics being a limiting case), it is obvious that in
  the absence of time dilation (lapse), shear and rotation, a geodesic
  must locally take the shortest spatial path, thus having vanishing projected curvature.}.

Notice also that whatever slicing we choose in a general spacetime,
we get a corresponding rescaled spacetime with a line
element of the form BlockDiag$[-1,g_{ij}(t,{\bf x})]$, where the motion of
free photons corresponds to null geodesics. In the rescaled spacetime the only degrees of freedom
are those of the spatial metric. So from this point of view, we can
always have a space that accounts for the behavior of photons. Indeed the
formalism of the new-straight curvature can be seen as a certain way
of describing {\it how} that space dictates the motion of geodesic photons.

\section{The generalized optical geometry in coordinates, assuming vanishing shear}
In order to get a better feeling for the properties of
spacetimes that admits a hypersurface-forming shearfree vector field,
it is instructive to carry out an analysis in
coordinates adapted to the chosen time slices and the corresponding
orthogonal congruence. In such coordinates, the rescaled metric takes the following form
\begin{eqnarray}\label{go}
\tilde{g}_{\mu\nu}=\left[
\begin{array}{llllll}
-1  & 0 \\
\ \ 0  & h_{ij}
\end{array}
\right]
.
\end{eqnarray}
In a spacetime of the form \eq{go} the only non-zero elements of the affine connection are
\begin{eqnarray}
\tilde{\Gamma}^0_{ij}&=&\frac{1}{2} \partial_t h_{ij} \\
\tilde{\Gamma}^k_{i0}&=& \frac{1}{2} h^{kl} \partial_t h_{li}=\tilde{\Gamma}^k_{0i} \label{hat} \\
\tilde{\Gamma}^k_{ij}&=& \frac{1}{2} h^{kl} \left( \partial_i h_{lj} + \partial_j h_{il} - \partial_l h_{ij} \right) \equiv \tilde{\gamma}^k_{ij}
.
\end{eqnarray}
Using these relations it is easy to evaluate the shear-expansion
tensor in the coordinates in question
\begin{eqnarray}
\tilde{\theta}_{ij}&=&\frac{1}{2} (\tilde{\nabla}_i \tilde{\eta}_j -
\tilde{\nabla}_j \tilde{\eta}_i  ) =... \nonumber  \\
&=&\frac{1}{2}\partial_t h_{ij} \label{kratt}
.
\end{eqnarray}
Here Latin indices run from 1 to 3.
The other components of $\tilde{\theta}_{\alpha \beta}$ are zero.
As mentioned before, the shear tensor vanishes if and only if
$[{\tilde{\theta}^\mu}{}_\alpha \tilde{t}^\alpha]_\perp=0$ for all
$\tilde{t}^\mu$. Hence, using \eq{kratt} (raising the first free
index to get ${\tilde{\theta}^i}{}_j=\frac{1}{2} h^{ik} \partial_t h_{kj}$),  necessary and sufficient conditions for the congruence shear to
vanish (in the coordinates in question) is
\begin{eqnarray}\label{extra1}
h^{ik} \partial_t h_{kj}\propto {\delta^i}_j 
.
\end{eqnarray}
Multiplying both sides by $h_{li}$ yields
\begin{eqnarray}\label{prop}
\partial_t h_{lj} \propto  h_{lj}
.
\end{eqnarray}
What this means is that when moving in time only, all the components of $h_{ij}$ must increase with
the same factor, for every fixed $\bf{x}$. The most general form for
$h_{ij}$ is then
\begin{eqnarray}
h_{ij}= e^{2 \Omega(t,\bf{x})} \bar{h}_{ij}({\bf x}) 
.
\end{eqnarray}
Here $\Omega$ is an arbitrary well behaved function of ${\bf x}$ and $t$, and $\bar{h}_{ij}({\bf x})$ is
independent of the time coordinate. Using this in \eq{go} yields 
\begin{equation} \label{finale}
\tilde{g}_{\mu\nu}=\left[
\begin{array}{lllll}
-1 \ \ \ \ \ \ \  \ \ \ 0  \ \ \ \ \ \ \  		\\[0mm]
\ \ 				\\[0mm]
\ \ 0 \ \ \ \ e^{2 \Omega(t,\bf{x})} \bar{h}_{ij}({\bf x})
\end{array}
\right]
.
\end{equation}
So, in coordinates adapted to the time slices and corresponding congruence, the line
element takes the above form if and only if the congruence is shearfree.
The conformally static spacetimes is a subset (set $\Omega=0$) of the
spacetimes described by conformal rescalings of \eq{finale}.

Incidentally one may use the above coordinate formalism to
verify the necessary and sufficient condition of
vanishing shear in order for (null) geodesics in the rescaled spacetime to
correspond to projected straight lines%
\footnote{The three spatial equations for a geodesic in the rescaled
  spacetime, in the coordinates in question, takes the form of
$\frac{d^2x^k}{d\tilde{\tau}^2} + \tilde{\gamma}^k_{i
    j}\frac{dx^i}{d\tilde{\tau}} \frac{dx^j}{d\tilde{\tau}} = -2 \tilde{\Gamma}^k_{i 0} 
\frac{dt}{d\tilde{\tau}} \frac{dx^i}{d\tilde{\tau}}$.
Since $\tilde{\gamma}^k_{i j}$ is the affine connection on the slice,
it is obvious that we must have 
$\tilde{\Gamma}^k_{i 0} \propto {\delta^k}_i$, in order for the projected trajectory to correspond to
a spatial geodesic. According to the above discussion, this
  holds if and only if the congruence shear vanishes.
}.

\subsection{The inertial force formalism in coordinates}\label{disk}
Inserting $h_{ij}=e^{2 \Omega(t,\bf{x})} \bar{h}_{ij}({\bf x})$ in
\eq{kratt} we find
\begin{eqnarray}
\tilde{\theta}_{ij}&=&(\partial_t \Omega)    h_{ij}
.
\end{eqnarray}
Using this relation in \eq{optisk} we readily find the optical inertial force formalism in coordinates adapted to the
congruence, assuming vanishing shear
\begin{eqnarray}\label{fin}
\frac{e^\Phi}{m} \left(\frac{F_\parallel}{\gamma}\tilde{t}^k + F_\perp \tilde{m}^k
\right)=h^{kl} \nabla_l \Phi + v  \tilde{t}^k \partial_t(\Phi+\Omega) +
\gamma^2 \frac{dv}{dt} \tilde{t}^k+\gamma^2 \frac{v^2}{\tilde{R}} \tilde{n}^k
.
\end{eqnarray}
Setting $\partial_t \Omega=0$, yields the inertial force
equation in the standard optical geometry for conformally static
spacetimes. Also setting $\partial_t \Phi=0$ yields the inertial
force equation in standard optical geometry for static spacetimes.

Multiplying \eq{fin} by $m$ and and shifting the first two terms
on the right hand side to the left,  we may identify two different
types of inertial forces as
\begin{eqnarray}\label{list}
\textnormal{Gravity}&:&-m h^{kl}\nabla_{l}\Phi\nonumber\\
\textnormal{Expansion}&:&-m(\partial_{t}\Phi+\partial_{t}\Omega)v^{k}\nonumber
.
\end{eqnarray}
The term 'Gravity' is introduced by analogy to the Newtonian sense of
gravity. From the point of view of general relativity, this term
would simply be called acceleration (referring to the congruence
acceleration). Strictly speaking, the term 'Expansion' refers to expansion in the
non-rescaled spacetime%
\footnote{
It is easy to show, for instance comparing \eq{fin} with \eq{much1c}
and using the formulas for how the kinematical invariants are
transformed as listed in \cite{rickinert}, for the particular case of
vanishing shear,
that we have $\partial_t(\Phi +\Omega)=e^\Phi \frac{1}{3}\theta$.}.
For positive $\partial_t(\Phi +\Omega)$
the term has the form of a viscous damping force although
for negative $\partial_t(\Phi +\Omega)$ it is rather a velocity proportional driving force.
The last two terms of \eq{fin} are a representation
of the motion (acceleration) relative to the reference congruence, and are not
regarded as inertial forces. For further discussion of these types of
interpretations, see \cite{rickinert}. Note however that precisely what we call an
inertial force is ambiguous up to factors of $\gamma$. We could for
instance multiply the entire equation, or just the parallel part of
the equation, by $\gamma$ and thus introduce
$\gamma$-factors in the inertial forces.  

\section{A note on given and received forces}\label{fnote}
In the formalism thus far presented, we have expressed forces in terms
of what is experienced by an observer comoving with the test particle
in question. These forces are in general different from the forces
needed to be {\it given} by the congruence observers, in order to make the test particle move as
specified by the curvature, curvature direction and the time
derivative of the speed. In \cite{rickinert} the relationship between
the given forces $F_{\script{c}\perp}$ and $F_{\script{c}\parallel}$
(where 'c' stands for congruence) and the received (comoving) forces
$F_{\perp}$ and $F_{\parallel}$ is derived 
\begin{eqnarray}
F_{\script{c}\perp}&=&\frac{F_\perp}{\gamma} \\  
F_{\script{c}\parallel}&=&F_\parallel 
\end{eqnarray}
These relations can be used to get the given forces in any of the
different formulations of this article (\eq{much1c},\eq{optisk} and
\eq{fin}). For instance we may consider a railway track in some static
geometry. The perpendicular force exerted by the track (as seen from the rest frame
of the track) on the train is given by $F_{\script{c}\perp}=\frac{F_\perp}{\gamma}$.
Note that if the track has vanishing optical curvature, this force (unlike the comoving force) has a velocity
dependence. Indeed it is easy to find \cite{intu} that there is no way to
  lay out a railway track (no curvature) such that the sideways {\it given}
  force is velocity independent.

\section{A note on uniqueness}\label{unique}
In the scheme of section \ref{news} (employing the novel curvature measure), we can
do optical geometry in a finite region around any point in any
spacetime. There are always more than one possible such geometry
(corresponding to different choices of slices). 

The optical geometry
as generalized in section \ref{projgen} (using the projected
curvature) is however more restrictive and
the standard optical geometry is more restrictive still. In
the coming two subsections we comment on the similarities and
differences between the standard and the 
generalized optical geometry of section \ref{projgen}.

\subsection{The standard optical geometry}
Standard optical geometry is defined for conformally static
spacetimes. These are defined as spacetimes that admits a timelike
hypersurface forming conformal Killing field. 
Mathematically this amounts to that
there must exist a scalar field $f$ and a vector field $\xi^\mu$ that
obeys%
\footnote{Assuming four dimensions, it follows from \eq{hupp} that
  $f=\frac{1}{2} \nabla_\alpha \xi^\alpha$. 
}
\begin{eqnarray}
\hspace{1.28cm}(\nabla_\mu \xi_\nu+\nabla_\mu \xi_\nu)&=&f g_{\mu\nu} \label{hupp}\\
{P_\mu}^\rho {P_\nu}^\sigma (\nabla_\sigma \xi_\rho -\nabla_\rho
\xi_\sigma)&=&0 \label{hit}
\end{eqnarray}
Here ${P_\mu}^\rho={\delta_\mu}^\rho + \frac{1}{-\xi^\sigma
  \xi_\sigma} \xi_\mu \xi^\rho$. The first equation is the conformal
  Killing equation. The second equation corresponds to setting
  $\omega_{\mu\nu}$ as defined in \eq{omdef}, to zero%
\footnote{Substitute $\eta^\mu$ by $\xi^\mu$, and modify the form
  of ${P_\mu}^\rho$ as was just shown.}.
For the particular case when $f=0$
  we have a Killing field rather than a conformal Killing field.
Given a field $\xi^\mu$ that obeys \eq{hupp} and \eq{hit}, we define $e^{2\Phi}=-g_{\alpha \beta} \xi^\alpha
\xi^\beta$. After a rescaling $\tilde{g}_{\mu\nu}=e^{-2\Phi}
g_{\mu\nu}$ we get our optical geometry.
Notice however that for any solution ($\xi^\mu,f$), we can form another
solution as ($\alpha \xi^\mu,\alpha f$) for some constant $\alpha$. Thus
the definition of $e^{2\Phi}$ is not unique. For simple cases, like a
Schwarzschild black hole, where we have asymptotic flatness, we can
however choose $\alpha$ so that $\xi^\mu$ is normalized at infinity.

There can however be a larger freedom still in $\xi$. For instance in
flat spacetime, in inertial coordinates, {\it any}
normalized timelike constant vector field satisfies the requirements
(although the optical geometry is flat for all choices). 
There are however also other, not quite so trivial, timelike
 hypersurface forming Killing fields for a flat spacetime. In particular 
there is a Killing field parallel to the four-velocities of the
points of a rigidly accelerating system (a so called Rindler
system\cite{rindler}). The associated optical geometry is here curved.
Note that this Killing field is not a global field however. 

In summary, there may
exist no standard optical geometry, and there may exist more than one
(non-trivially related) standard optical geometry, depending on the spacetime
in question. 

\subsection{The generalized optical geometry for the projected curvature}
In the generalized optical geometry of section \ref{projgen} (using
the projected curvature), the fundamental equation (requirement) is
not really an equation for a vector field, but for a congruence of
worldlines. There has to exist a non-rotating, non-shearing congruence
for this type of generalized optical geometry to work. This is however
equivalent to the existence of a rotationfree and shearfree
vector field $\zeta^\mu$ (not necessarily normalized) that obeys
\begin{eqnarray}
{P_\mu}^\rho {P_\nu}^\sigma  (\nabla_\rho \zeta_\sigma+\nabla_\sigma
\zeta_\rho)=f P_{\mu \nu} \label{hh}\\
{P_\mu}^\rho {P_\nu}^\sigma (\nabla_\sigma \zeta_\rho -\nabla_\rho
\zeta_\sigma)=0 \label{hhh}
\end{eqnarray}
Here ${P_\mu}^\rho={\delta_\mu}^\rho + \frac{1}{-\zeta^\sigma
  \zeta_\sigma} \zeta_\mu \zeta^\rho$ and again $f$ is some scalar
  function%
\footnote{In four dimensions it follows that 
$f=\frac{2}{3} P^{\alpha \beta} \nabla_\alpha \zeta_\beta$. }.
The first equation corresponds to setting $\sigma_{\mu \nu}$ as
  defined in \eq{sidef} to zero%
\footnote{Substitute $\zeta^\mu$ by $\xi^\mu$, and modify the form
  of ${P_\mu}^\rho$ as was just shown. Note that
 when applying \eq{sidef} to a non-normalized vector
 field, the appropriate $\theta$ (corresponding to $\frac{3}{2}f$) is given by  $\theta=P^{\alpha \beta}\nabla_\alpha \zeta_\beta$.}.
The second equation corresponds to setting
  $\omega_{\mu\nu}$ as defined in \eq{omdef} to zero%
\footnote{Substitute $\zeta^\mu$ by $\xi^\mu$, and modify
  the form of ${P_\mu}^\rho$ as was just shown.}.
Expressed in this form we note that the only difference from the
standard optical geometry equations (substituting $\zeta^\mu$ with
$\xi^\mu$) lies in the projection operators in \eq{hh}. Indeed \eq{hh}
  is the projected version of \eq{hupp}.
This means that the standard optical geometry equations are more restrictive than
the latter two equations. Any field that satisfies \eq{hupp}
and \eq{hit}, will also satisfy \eq{hh} and \eq{hhh},
  but the converse is not generally true.

So, given a field $\zeta^\mu$ that satisfies \eq{hh} and \eq{hhh} the
generalized optical geometry exists. We cannot however in general find the
rescaling parameter (modulo a single constant) from the norm of
$\zeta^\mu$%
\footnote{In fact \eq{hh} and
\eq{hhh} are independent of the norm of $\zeta^\mu$ in the sense that
if $\zeta^\mu$ solves \eq{hh} and
\eq{hhh} so will $h \zeta^\mu$ where $h$ is some arbitrary
function. This is contrary to \eq{hupp} of the preceeding section, 
where given a field $\xi^\mu$ that solves \eq{hupp}, a field $h
\xi^\mu$ would solve \eq{hupp} if and only if $h$ was constant.}. To find the
rescaling we would instead consider the slices orthogonal to the
field and assign a continuous parameter $t$ to these slices. We would
then make a rescaling 
so that
$\tilde{g}^{\alpha \beta} \nabla_\alpha t \nabla_\beta t=-1$.  
We may understand that the freedom in the labeling $t$ of the time slices, gives a freedom
of a scale factor (as a
function of $t$) for the whole (spatial) optical geometry. For any standard optical geometry there
are thus a multitude of generalized optical geometries, for the same
reference congruence. As we will demonstrate in a forthcoming paper
\cite{application} one can also have a standard optical geometry
connected to a certain reference congruence, and in the same region of spacetime
have a generalized optical geometry for a different congruence. 

More importantly however, there are cases where there exists no standard optical
geometry, but where the generalized optical geometry exists. In the
companion paper \cite{application} we illustrate that there exists a generalized
optical geometry (within a finite region of any spacetime point) for
any spherically symmetric spacetime. Indeed there are infinitely many
(in-falling) non-shearing reference congruences for this case. In particular we show that there
exists a generalized optical geometry across the horizon of a Schwarzschild black hole
(unlike in the standard optical geometry).

The equations \eq{hh} and \eq{hhh} are here mainly included for
comparison with the standard optical geometry equations. In the
companion paper \cite{application}, we in fact find it more convenient to
work in coordinates, starting from \eq{finale}.

\section{Summary and conclusion}
A generalization of the optical geometry has been proposed prior to
this paper \cite{ANWsta} using a different philosophy, but see \cite{jantzen2} for criticism.  
The optical geometry as presented in this article can be done at different levels.
\begin{itemize}

\item
In any spacetime we can produce an optical geometry where photons move with unit speed. 
Using the new sense of curvature \cite{rickinert}, a geodesic photon follows a
spatially straight line and the sideways force on a massive particle
following a straight line is independent of the velocity.

\item
In any spacetime that admits a hypersurface orthogonal shearfree
congruence of worldlines, we can produce an optical geometry where
photons move with constant speed. Here the projected curvature and the
new-straight curvature coincide and therefore either one can be used to define what is
spatially straight.  A geodesic photon follows a
spatially straight line and the sideways force on a massive particle
following a straight line is independent of the velocity. Furthermore, a
gyroscope initially pointing in the direction of motion, following
a spatially straight line, will not precess (independently of the velocity) relative to the forward
direction. If the gyroscope follows a spatially curved line, it obeys
a very simple law of three-dimensional precession \eq{ratt2}.

\item
In a conformally static spacetime (choosing the preferred congruence)
we have the same features of the optical geometry as outlined in the
preceding point. Here the optical geometry is static. 
Furthermore, as demonstrated in \cite{Maxwell}, Maxwell's equations
take a simple form written in terms of the optical metric.

\end{itemize}
In any generalization of a theory, there are in general several
possibilities. Indeed if we consider the standard optical geometry, in
conformally static spacetimes, we may use either the projected or the
new-straight curvature (they are identical here). Generalizing to more
complicated spacetimes we however get two different theories (at least
algebraically) depending on what curvature measure we are adopting. 
One may certainly consider other ways of defining an optical geometry. For
instance, one might consider relaxing the spatial
geodesic requirement of photons and replace it with some preferably
simple law (indeed that is what we get if we consider the projected
curvature when there is shear, see \eq{much1c}). 

Apart from extending the set of spacetimes where one can introduce
optical geometry, this article also aims to present a solid
inertial force formalism both for the generalized and the standard
optical geometry (again see \cite{jantzen2} for criticism of
previous works).

As regards applications of the generalized optical geometry we refer
to a companion paper \cite{application} where a non-static
(but shearfree) congruence is employed to do
optical geometry across the horizon of a static black hole.

\appendix

\section{The Fermi-Walker equation for the gyroscope spin vector in
the rescaled spacetime}\label{fermiapp}
Consider a rescaled spacetime
$\tilde{g}_{\mu\nu}=e^{-2\Phi}g_{\mu\nu}$. Let $k^\mu$ be a general
vector defined along a worldline of four-velocity $v^\mu$. Introducing
$\tilde{v}^\mu=e^\Phi v^\mu$ and $\tilde{k}^\mu=e^\Phi k^\mu$ one may
show \cite{rickinert} that the relation between the rescaled and
the non-rescaled covariant derivative of the vector $k^\mu$ is given by  
\begin{eqnarray}\label{dero2}
\frac{\tilde{D} \tilde{k}^\mu}{\tilde{D}\tilde{\tau}}=e^{2\Phi}\left( \frac{Dk^\mu}{D\tau} -(v^\mu k^\rho -
v^\alpha k_\alpha   g^{\mu \rho}) \nabla_\rho \Phi \right) \label{dero}
.
\end{eqnarray}
In particular, for $k^\mu=v^\mu$, we get the transformation
of the four-acceleration
\begin{eqnarray}\label{fouracc2}
\frac{\tilde{D} \tilde{v}^\mu}{\tilde{D}\tilde{\tau}}&=&
e^{2\Phi}\left( \frac{Dv^\mu}{D\tau} -(v^\mu v^\rho + g^{\mu \rho} ) \nabla_\rho \Phi \right)
.
\end{eqnarray}
The Fermi-Walker equation for a gyroscope of spin vector
$S^\mu$ is given by
\begin{eqnarray}
\frac{D S^\mu}{D\tau} = v^\mu S_\alpha \frac{D v^\alpha}{D\tau}
.
\end{eqnarray}
Using \eq{dero2}, substituting $k^\mu \rightarrow S^\mu$,  and
\eq{fouracc2} together with the $S_\alpha v^\alpha=0$ we readily find 
\begin{eqnarray}
\frac{\tilde{D} \tilde{S}^\mu}{\tilde{D} \tilde{\tau}} = \tilde{v}^\mu \tilde{S}_\alpha\frac{\tilde{D} \tilde{v}^\alpha}{\tilde{D} \tilde{\tau}} 
.
\end{eqnarray}
Thus a spin vector that is Fermi-Walker transported relative to the
non-rescaled spacetime is Fermi-Walker transported also relative to the
rescaled spacetime if we just rescale the spin vector itself. The
converse obviously holds also.

\section{Shearfree congruences}\label{shearl}
Assuming that $[{\tilde{\theta}^\mu}{}_\beta \tilde{t}^\beta]_\perp=0$,
for all directions $\tilde{t}^\mu$ we have
\begin{eqnarray}\label{kol}
 {\tilde{\theta}^\mu}{}_\beta \tilde{t}^\beta \propto \tilde{t}^\mu
.
\end{eqnarray}
Knowing that ${\tilde{\theta}^\mu}{}_\beta =
{\tilde{\sigma}^\mu}{}_\beta + \frac{\tilde{\theta}}{3} {\tilde{P}^\mu}{}_\beta$ we see
that \eq{kol} is equivalent to ${\tilde{\sigma}^\mu}{}_\beta
\tilde{t}^\beta \propto \tilde{t}^\mu$. We know that ${\tilde{\sigma}^\mu}{}_\beta
\tilde{\eta}^\beta=0$. Since $\tilde{\sigma}_{\mu \nu}$ is a symmetric tensor it follows that in
coordinates adapted to the congruence, only the spatial part of
$\tilde{\sigma}_{\mu \nu}$ is nonzero. Also, for \eq{kol} to hold for arbitrary
spatial directions $\tilde{t}^i$, we must have $\tilde{\sigma}^i{}_j
\propto \delta^i{}_j$. Knowing also that the trace
$\tilde{\sigma}^\alpha{}_\alpha$ always vanishes, it follows that
$\tilde{\sigma}^\mu{}_\nu$ must vanish entirely. This in turn is true
if and only if the non-rescaled shear tensor vanishes since 
$\tilde{\sigma}_{\mu \nu}=e^{-\Phi} \sigma_{\mu \nu}$.
\\
\\
{\bf References}
\\

\end{document}